\documentclass[12pt]{iopart}

\usepackage{amssymb}

\usepackage{iopams}  

\usepackage{cite}
\usepackage{graphicx}
\usepackage{multirow}
\usepackage{bm}

\usepackage{color}

\begin{document}

\title{Noncyclic geometric phase in counting statistics and its role as an excess contribution}


\author{Jun Ohkubo}

\address{
Graduate School of Informatics, Kyoto University,\\
36-1, Yoshida Hon-machi, Sakyo-ku, Kyoto-shi, Kyoto 606-8501, Japan
}
\ead{ohkubo@i.kyoto-u.ac.jp}
\begin{abstract}
We propose an application of fiber bundles to counting statistics.
The framework of the fiber bundles gives a splitting of a cumulant generating function
for current in a stochastic process,
i.e., contributions from the dynamical phase and the geometric phase.
We will show that the introduced noncyclic geometric phase is related
to a kind of excess contributions, which have been investigated a lot in nonequilibrium physics.
Using a specific nonequilibrium model,
the characteristics of the noncyclic geometric phase are discussed;
especially, we reveal differences between a geometric contribution for the entropy production 
and the `excess entropy production' which has been used to discuss the second law of steady state thermodynamics.
\end{abstract}

\maketitle

\section{Introduction}

Studies of nonequilibrium systems have been focused a lot mainly in the last decade.
One of the main topics is the transitions between stationary states.
During the transitions, a system causes a kind of flows or currents of physical quantities.
For example, as for a transition between two equilibrium states,
the change of thermodynamic quantities can be calculated using equilibrium thermodynamics.
However, different from the equilibrium systems,
the nonequilibrium systems have stationary flows or currents, which stem from their nonequilibrium nature
even if we do not consider any transitions between stationary states.
In order to investigate such nonequilibrium systems,
a concept of `excess' contributions play an important role
\cite{Oono1998,Hatano2001,Speck2005,Esposito2007,Komatsu2008,Komatsu2008a,Komatsu2009,Esposito2010}.
That is, if we change the system using external forces,
the nonequilibrium systems can generate extra currents caused by the change of states,
in addition to the contributions from the steady currents.
For example, an excess heat and an excess entropy production have been introduced 
in order to discuss the second law of steady state thermodynamics
\cite{Hatano2001,Esposito2007}.
However, when we consider a physical quantity related to nonequilibrium states,
the definition of the excess contribution of the physical quantity is not obvious
and there may be many possibilities for the excess contribution.

Recently, the geometric phase concept has been used to investigate flows or currents in stochastic processes.
For cyclic perturbations, the geometric phase concept has given successful results 
\cite{Sinitsyn2007,Sinitsyn2007a,Ohkubo2008,Ohkubo2008a,Sinitsyn2008,Sinitsyn2009}.
However, in order to study the excess quantities for general cases,
it is needed to treat noncyclic frameworks;
for example, we consider a sudden change of the transition rates of a stochastic system,
and then the system does not return to the initial state.
When the change of the transition rates is very slow (so-called adiabatic cases),
a path-integral formulation is available and 
a quantity called a noncyclic geometric phase has been introduced.
To our knowledge, at least two different definitions have been proposed \cite{Sinitsyn2010,Sagawa2011}.
As for the nonadiabatic cases, 
a definition of the geometric phase has been introduced from a mathematical viewpoint \cite{Ohkubo2010}.
However, it has not been clear whether the proposed geometric phase in \cite{Ohkubo2010} has physical meanings or not.

In the present paper,
a new definition of the geometric phase is proposed.
Although the proposed framework is similar to the previous work \cite{Ohkubo2010},
it gives the following consequence:
the contribution from the geometric phase has the characteristics of the excess contributions.
In addition, the framework can be used to discuss arbitrary physical quantities related to nonequilibrium current;
in the present paper, we demonstrate the framework by using the entropy productions in nonequilibrium states.

The present paper is constructed as follows.
In section 2 we explain the problem settings of the counting statistics.
Section 3 gives the main result of the present paper;
a new definition of the geometric phase is introduced,
and we discuss the geometric nature based on the fiber bundles.
In section 4, we give some examples of the proposed geometric phase,
and confirm that the proposed geometric phase
has the characteristics of the excess contribution.
In addition, the entropy production is discussed,
and we show differences between the contribution from the noncyclic geometric phase and
the `excess entropy production' studied previously \cite{Esposito2007}.
Section 5 gives some concluding remarks.

\section{Problem settings}

We consider a stochastic process with finite discrete states; the number of states is $S$.
The dynamics of the stochastic process is described by the following master equation:
\begin{eqnarray}
\frac{\rmd}{\rmd t} | p(t) \rangle = K(t) | p(t) \rangle,
\label{eq_master}
\end{eqnarray}
where $|p(t)\rangle = (p_1(t), p_2(t), \dots, p_S(t))^\mathrm{T}$ is a vector for the probability distribution
of the system and 
$K(t) = \{ K_{ss'}(t) \}$ the transition matrix;
the element $K_{ss'}(t)$ is related to the transition rate for the state change $s' \to s$ at time $t$.
For generality, we consider that the state change $s' \to s$ 
is caused through several ways (see the example in section 4).
For each state change $s' \to s$,
we consider the transition rates 
$w_{ss'}^{1}(t), w_{ss'}^{2}(t), \dots, w_{ss'}^{V_{ss'}}(t)$;
totally $V_{ss'}$ ways of transitions exist for the state change $s' \to s$.
In addition, we set $V_{ss} = 1$ for all $s$ and
$w_{ss}^{1}(t) = 0$.
Using the transition rates $\{w_{ss'}^{\nu}(t)\}$,
the elements of the transition matrix $K(t)$ are written as follows:
\begin{eqnarray}
K_{ss'}(t) \equiv \sum_{\nu = 1}^{V_{ss'}} w_{ss'}^{\nu}(t) - \lambda_{s}(t) \delta_{s,s'},
\label{eq_definition_of_transition_matrix}
\end{eqnarray}
where $\delta_{s,s'}$ is the Kronecker's delta and
\begin{eqnarray}
\lambda_{s}(t) = \sum_{s'=1}^{S} \sum_{\nu=1}^{V_{s's}}  w_{s's}^{\nu}(t).
\end{eqnarray}

We next consider `currents' in the stochastic process.
One example of the currents is the number of the occurrence of specific transitions 
(for example, see \cite{Sinitsyn2007}),
and another one is the entropy production (for example, see \cite{Sagawa2011}).
Here, we consider a quantity $\alpha_{ss'}^{\nu}$
defined for each state change $s' \to s$ and the way of the transition;
if we have one state change $s' \to s$ by the $\nu$-th transition way,
we count a quantity $\alpha_{ss'}^{\nu}$.
Our aim is to calculate the statistics of an accumulated quantity.
In order to define the accumulated quantity,
we need information about a trajectory and the transition ways, as follows.
We fix a time interval $[0,T]$ and consider a trajectory of the state change,
$[s(t)]_{t=0}^T$, where $s(t) \in S$.
In other words, 
the trajectory is characterized by jump events at discrete times $(t_1, t_2, \dots, t_{n-1})$,
and expressed as 
$s(t) = s_i$ for $t_{i-1} \le t < t_i$ with $t_0 = 0$ and $t_n = T$.
In addition, the jump event at time $t_i$
is specified by the $\nu_i$-th way of transitions for the state change $s' \to s$.
Then, the accumulated quantity is defined as
\begin{eqnarray}
A(T) = \sum_{i=1}^{n-1} \alpha_{s_{i+1} s_{i}}^{\nu_i},
\label{eq_accumulated_quantity}
\end{eqnarray}
where $\nu_i \in \{1,\dots,V_{s_{i+1} s_{i}}\}$.
The generating function for the accumulated quantity $A(T)$ is given by
\begin{eqnarray}
F(\chi,T) \equiv \left\langle \rme^{\chi A(T)} \right\rangle,
\label{eq_original_generating_function}
\end{eqnarray}
where the average $\langle \cdot \rangle$ is taken over realizations of 
trajectories with an initial distribution $|p(0)\rangle$.
Derivatives of $F(\chi,T)$ with respect to $\chi$ give
various information about the statistics of the accumulated quantity.

The problem here is: How should we calculate \eref{eq_original_generating_function}?
One may consider that it is difficult to take the average over the trajectories.
In order to evaluate \eref{eq_original_generating_function},
the framework of the counting statistics has been developed well 
and used in various context (for example, see \cite{Gopich2006}).
The consequences of the counting statistics give us the following simple way to evaluate the generating function.
For details of the derivations, see \cite{Garrahan2009}, for example.

We here consider the following matrix;
\begin{eqnarray}
H_{ss'}(\chi,t) = \sum_{\nu = 1}^{V_{ss'}} w_{ss'}^{\nu}(t) \exp\left( \chi \alpha_{ss'}^{\nu}\right) 
- \lambda_{s}(t) \delta_{s,s'}.
\label{eq_modified_transition_matrix}
\end{eqnarray}
Note that the matrix $H(\chi,t) = \{H_{ss'}(\chi,t)\}$ is not a `probability' transition matrix.
Using the matrix $H(\chi,t)$,
the following time-evolution equation is considered:
\begin{eqnarray}
\frac{\rmd}{\rmd t} | \psi(\chi,t) \rangle
= H(\chi,t) | \psi(\chi,t) \rangle
\label{eq_time_evolution_for_psi_ket}.
\end{eqnarray}
Again, note that the vector $|\psi(\chi,t)\rangle$ is the probability distribution;
using a vector $\langle 1 | \equiv (1,1,\dots,1)$, 
we have $\langle 1 | \psi(\chi,t) \rangle \neq 1$ in general.

The remarkable consequence of the counting statistics is as follows:
Using the solution of \eref{eq_time_evolution_for_psi_ket} at time $T$
with an initial condition $|\psi(\chi,t=0)\rangle = | p(0) \rangle$,
the generating function in \eref{eq_original_generating_function}
is given by
\begin{eqnarray}
F(\chi,T) = \langle 1 | \psi(\chi,T)\rangle.
\label{eq_generating_function_by_ket}
\end{eqnarray}

\section{A new geometric phase and fiber bundles}

All statistics of $A(T)$ defined by \eref{eq_accumulated_quantity} 
are given by the generating function \eref{eq_generating_function_by_ket},
and here we rewrite the framework 
from the viewpoint of the fiber bundles \cite{Bohm_book,Nakahara_book}.
The following discussions give a splitting of the generating function into two parts,
and we will show that one of them has the characteristics of the excess contribution.
In the following subsections, firstly, a definition of the geometric phase is given,
and we confirm that the introduced phase has actually a geometric nature.
After that, we comment on its characteristics as the excess contribution.


\subsection{A new definition of the geometric phase}

In addition to the `ket' vector $| \psi(\chi,t) \rangle$ introduced in section 2,
a `bra' vector $\langle \psi(\chi,t)|$ is introduced
and its time-evolution is defined as follows:
\begin{eqnarray}
\frac{\rmd}{\rmd t} \langle \psi(\chi,t) |
= \langle \psi(\chi,t) | H(\chi,t),
\label{eq_time_evolution_for_psi_bra}
\end{eqnarray}
where the initial condition is $\langle \psi(\chi,t=0)| = \langle 1 |$.
Note that $-H(\chi,t)$ has been used to define the time-evolution in \cite{Ohkubo2010},
and the difference of the time-evolution operator is important;
the new definition gives characteristics of the excess contribution, as discussed later.
Additionally, we note that 
$\langle \psi(\chi,t)| \neq ( | \psi(\chi,t)\rangle)^\mathrm{T}$ 
in general because of the non-Hermite property of $H(\chi,t)$.
(Of course, the initial condition is also different from that of $| \psi(\chi,t)\rangle$.)
Using the bra vector $\langle \psi(\chi,t)|$ and 
the ket vector $| \psi(\chi,t)\rangle$,
the generating function $F(\chi,T)$ is expressed as
\begin{eqnarray}
F(\chi,T) = \langle \psi(\chi,0) | \psi(\chi,T) \rangle.
\end{eqnarray}

The inner product of the bra vector $\langle \psi(\chi,t)|$ and 
the ket vector $| \psi(\chi,t)\rangle$ is not conserved in general
because
$\frac{\rmd}{\rmd t} \left(\langle \psi(\chi,t) | \psi(\chi,t) \rangle \right) 
= 2 \langle \psi(\chi,t) | H(\chi,t) |\psi(\chi,t) \rangle$.
Instead of $\langle \psi(\chi,t)|$ and $| \psi(\chi,t)\rangle$,
we introduce the corresponding vectors
whose inner product is explicitly conserved.
That is, defining a quantity $\delta(\chi,t)$ as
\begin{eqnarray}
\delta(\chi,t') \equiv 
\frac{\langle \psi(\chi,t') | H(\chi,t') | \psi(\chi,t') \rangle}
{\langle \psi(\chi,t') | \psi(\chi,t') \rangle},
\label{eq_original_definition_of_delta}
\end{eqnarray}
the following new bra and ket vectors are introduced:
\begin{eqnarray}
| \phi(\chi,t) \rangle \equiv \exp\left( - \int_0^t \delta(\chi,t') \rmd t' \right) | \psi(\chi,t) \rangle, \\
\langle \varphi(\chi,t) | \equiv \exp\left( - \int_0^t \delta(\chi,t') \rmd t' \right) \langle \psi(\chi,t) |.
\end{eqnarray}
Hence, the time-evolutions of these vectors are given as
\begin{eqnarray}
\frac{\rmd}{\rmd t} | \phi(\chi,t) \rangle
= \left[ H(\chi,t) - \delta(\chi,t) \right] | \phi(\chi,t) \rangle,
\label{eq_time_evolution_for_phi_ket}
\end{eqnarray}
\begin{eqnarray}
\frac{\rmd}{\rmd t} \langle \varphi(\chi,t) |
= \langle \varphi(\chi,t) | \left[ H(\chi,t) - \delta(\chi,t) \right].
\label{eq_time_evolution_for_phi_bra}
\end{eqnarray}
Note that 
the inner product is conserved because 
\begin{eqnarray}
\frac{\rmd}{\rmd t} \left( \langle \varphi(\chi,t) | \phi(\chi,t) \rangle \right) = 0,
\end{eqnarray}
and from the initial conditions, we have
\begin{eqnarray}
\langle \varphi(\chi,t) | \phi(\chi,t) \rangle = 1.
\end{eqnarray}

Using the new bra and ket vectors $\langle \varphi(\chi,t)|$ and $| \phi(\chi,t)\rangle$,
the generating function is rewritten as
\begin{eqnarray}
F(\chi,T)
&= \langle \varphi(\chi,0) | \exp\left[\int_0^T \delta(\chi,t) \rmd t \right] | \phi(\chi,T) \rangle 
\nonumber \\
&= \exp\left[\int_0^T \delta(\chi,t) \rmd t  +  
\ln \langle \varphi(\chi,0) | \phi(\chi,T) \rangle \right].
\end{eqnarray}
Hence, when we write the cumulant generating function as $\mu(\chi,t) \equiv \log F(\chi,t)$,
the cumulant generating function is adequately divided into two parts:
\begin{eqnarray}
\mu(\chi,T) = \int_0^T \delta(\chi,t) \rmd t + 
\ln \langle \varphi(\chi,0) | \phi(\chi,T) \rangle.
\label{eq_division_of_cumulant}
\end{eqnarray}
The first term in the right hand side in \eref{eq_division_of_cumulant}
is called the dynamical phase,
and the second term corresponds to the geometric phase.
Although someone feels the word `phase' is strange in this case because 
they are not imaginary numbers,
we employ the conventional word `phase' here.

While solutions of \eref{eq_time_evolution_for_psi_ket} 
is enough to evaluate the generating function $F(\chi,T)$,
only the solutions of 
\eref{eq_time_evolution_for_phi_ket} and \eref{eq_time_evolution_for_phi_bra} are needed
in order to calculate the geometric phase;
the geometric phase is immediately given from $| \phi(\chi,T)\rangle$,
and the dynamical phase $\int \delta(\chi,t) \rmd t$
is also evaluated directly from $\langle \varphi(\chi,t)|$
and $| \phi(\chi,t)\rangle$
with some modifications of \eref{eq_original_definition_of_delta}.
In order to evaluate other quantities, such as the average, deviations, and so on,
evaluations of other optional time-evolution equations are useful;
see section 4.2, for example.

\subsection{Fiber bundles}

The definitions and constructions in section 3.1 are enough to calculate 
the `geometric' phase,
but it is still unclear why the defined quantity is called the `geometric' phase,
at this stage.
In this subsection, we discuss the mathematical aspects of the defined `geometric' phase,
which makes the meanings of `geometric' clear.

We consider the principal fiber bundle $P$ whose fiber $F$ is given by 
$\mathbb{R}_{+} = \{x \in \mathbb{R}| x > 0\}$.
The base space $M$ is constructed as follows:
$P$ is constructed by the set of all states $\{| \phi(\chi,t)\rangle\}$,
and we introduce the equivalent relation $\sim$ as a kind of proportional relation;
if $| \phi(\chi,t) \rangle = c |\phi' (\chi,t') \rangle$ with $c > 0$,
$| \phi(\chi,t) \rangle$ and $|\phi' (\chi,t') \rangle$
are considered as equivalent ones.
We therefore have the base space $M = P/\sim$.
Using a natural projection map $\pi: P \to M$,
the construction of the principal fiber bundle is finished.
We also consider its dual constructed by $\{\langle \varphi(\chi,t)|\}$.

The following connection is introduced for $P$.
Let $| \phi(\chi,s) \rangle$ be a curve in $P$,
and write the tangent vector to this curve as $| u \rangle = \frac{\rmd}{\rmd s} | \phi (\chi,s) \rangle$.
Then, using a dual curve $\langle \varphi(\chi,s)|$, the connection is defined as
\begin{eqnarray}
A(\chi,s) = \frac{\langle \varphi(\chi,s) | u (\chi,s) \rangle}{\langle \varphi(\chi,s) | \phi (\chi,s) \rangle}.
\label{eq_definition_of_connection}
\end{eqnarray}
For the dual bundle,
we introduce a slightly different connection as follows:
\begin{eqnarray}
A^{\mathrm{d}}(\chi,s) = \frac{\langle v(\chi,s) | \phi (\chi,s) \rangle}{\langle \varphi(\chi,s) | \phi (\chi,s) \rangle},
\end{eqnarray}
where $\langle v | = \frac{\rmd}{\rmd s} \langle \varphi (\chi,s) |$.
Note that $\langle \varphi(\chi,s) | \phi (\chi,s) \rangle \neq 1$ in general.
In the present paper, we treat only cases with conserving inner products;
$\langle \varphi(\chi,s) | \phi (\chi,s) \rangle$ is constant.
That is, when we have a curve $|\phi(\chi,s)\rangle$ in $P$,
the dual curve $\langle \varphi(\chi,s)|$ is selected as 
$\frac{\rmd}{\rmd s} (\langle \varphi(\chi,s) | \phi (\chi,s) \rangle) = 0$.
Actually, the time-evolutions in
\eref{eq_time_evolution_for_phi_ket} and \eref{eq_time_evolution_for_phi_bra}
conserve the inner product,
and geodesic curves discussed later are assumed to have this property.
Hence, we have $A(\chi,s) = - A^{\mathrm{d}}(\chi,s)$.
Additionally, in the time-evolution of \eref{eq_time_evolution_for_phi_ket},
the connection is always zero because
\begin{eqnarray}
\langle \varphi(\chi,t) | \frac{\rmd}{\rmd t} | \phi(\chi,t) \rangle 
= 0.
\end{eqnarray}
Furthermore, the gauge transformation is given as
\begin{eqnarray}
&|\phi(\chi,s)\rangle \to e^{\beta(s)}|\phi(\chi,s)\rangle, \qquad
\langle \varphi(\chi,s) | \to \langle \varphi(\chi,s)| e^{-\beta(s)}, \nonumber \\
&A(\chi,s) \to A(\chi,s) + \frac{\rmd \beta(s)}{\rmd s}, 
\end{eqnarray}
where $\beta(s) \in \mathbb{R}$.

In the fiber bundles,
a `geometric' phase
is usually determined only by the shape of the trajectory of the time-evolution.
That is, for a closed curve in $P$,
the following geometric phase is introduced:
\begin{eqnarray}
\gamma(\chi,T) = \oint A(\chi,s) \rmd s.
\label{eq_def_of_gamma_closed_curve}
\end{eqnarray}
Although one may consider that the integral defined by the closed curve is meaningful
only when we consider a cyclic evolution for \eref{eq_time_evolution_for_phi_ket},
the following construction makes the integral \eref{eq_def_of_gamma_closed_curve} possible
even for noncyclic time-evolution cases.
The initial vector $| \phi(\chi,0)\rangle$ and the final vector 
$|\phi (\chi,T) \rangle$ does not make a closed curve in general cases;
there is only a one-way path from $| \phi(\chi,0)\rangle$
to $|\phi (\chi,T) \rangle$ obeying the time-evolution \eref{eq_time_evolution_for_phi_ket}.
In order to make a closed curve from the one-way path,
a geodesic curve is useful;
there are many possibilities to make a closed path,
but the geodesic curve guarantees 
the gauge invariant property of the defined quantity.
The basic framework has been written in \cite{Ohkubo2010,Samuel1988},
and we explain the details of the usage of the geodesic curve in Appendix.
The important point here is that
the integral on the closed curve is given as follows:
\begin{eqnarray}
\fl
\oint A(\chi,s) \rmd s &= 
\int_\textrm{\scriptsize (time-evolution)} A(\chi,s) \rmd s 
+ \int_\textrm{\scriptsize (geodesic curve)} A(\chi,s') \rmd s' 
\label{eq_integral_of_A} \\
\fl &= \ln \langle \varphi(\chi,0) | \phi(\chi,T) \rangle.
\label{eq_integral_of_A_result}
\end{eqnarray}
The first term corresponds to the contribution from the actual time-evolution
from $|\phi(\chi,0)\rangle$ to $|\phi(\chi,T)\rangle$,
and the second term is the contribution from the geodesic curve
connecting $|\phi(\chi,T)\rangle$ and $|\phi(\chi,0)\rangle$.
Note that the first term in \eref{eq_integral_of_A} vanishes because
the connection defined by \eref{eq_definition_of_connection} is zero
on the time-evolution in \eref{eq_time_evolution_for_phi_ket} and
\eref{eq_time_evolution_for_phi_bra};
only the second term contributes and gives \eref{eq_integral_of_A_result}.
Although $A(\chi,s)$ in the first term in \eref{eq_integral_of_A} becomes non-zero
when one considers a gauge transformation,
the geometric quantity $\gamma(\chi,T)$ is gauge invariant.

Note that \eref{eq_integral_of_A_result}
gives the geometric phase in \eref{eq_division_of_cumulant} adequately.
Hence, we confirm that the framework of the fiber bundles
gives the natural splitting of the cumulant generating function.
In addition, 
while the geodesic curve is needed to define the integral of the connection $A(\chi,s)$,
there is no need to calculate the geodesic curve explicitly;
the solutions of \eref{eq_time_evolution_for_phi_ket} and \eref{eq_time_evolution_for_phi_bra}
is enough to calculate various quantities, as discussed before.

\subsection{Characteristics as excess contributions}

Excess contributions of some physical quantities are important
to discuss nonequilibrium states, as introduced in section 1.
In order to explain the excess contributions,
as an example, we here consider an entropy production \cite{Esposito2007}.
The entropy production depends on trajectories of the system,
and in order to evaluate the entropy production,
the quantity $\alpha_{s s'}^{\nu}$ in section 2
is defined as
\begin{eqnarray}
\alpha_{ss'}^{\nu} = \ln \frac{w_{ss'}^{\nu}}{w_{s's}^{\nu}} 
\label{eq_def_EP}
\end{eqnarray}
if $w_{ss'}^{\nu} \neq 0$ and $w_{s's}^{\nu} \neq 0$.
The accumulated entropy production is called the reservoir entropy production $S_\mathrm{r}$ \cite{Esposito2007}.
In addition, in \cite{Esposito2007},
it has been shown that the reservoir entropy production $S_\mathrm{r}$
can be divided into two parts as follows:
\begin{eqnarray}
S_\mathrm{r} = S_\mathrm{ex} + S_\mathrm{hk},
\end{eqnarray}
where $S_\mathrm{ex}$ and $S_\mathrm{hk}$ are called the excess entropy production and
the house keeping entropy production, respectively.
The house keeping entropy production corresponds to steadily generated entropy
in steady states, and the excess entropy is intrinsically related
to transitions between nonequilibrium steady states.
The excess entropy is defined as the integrals of the following time evolution equation:
\begin{eqnarray}
\frac{\rmd}{\rmd t} S_\mathrm{ex} = 
- \sum_{s, s'} \left( \sum_{\nu} w_{ss'}^{\nu}(t) \right) p_{s'}(t)
\ln \frac{p_{s'}^\mathrm{st}(t)}{p_{s}^\mathrm{st}(t)},
\label{eq_excess_EP}
\end{eqnarray}
where $p_{s}^\mathrm{st}(t)$ is the stationary probability distribution
for the specific transition rates $\{w_{ss'}^{\nu} (t) \}$ at time $t$.
The house keeping entropy production, $S_\mathrm{hk}$,
can be evaluated when we know $S_\mathrm{r}$ and $S_\mathrm{ex}$.
When the detailed balance condition is satisfied, 
we have $S_\mathrm{hk} = 0$;
for nonequilibrium states, $S_\mathrm{hk} \ge 0$.
In contrast, $S_\mathrm{ex}$ is zero when the system remains in a nonequilibrium states;
when we consider time-dependent transition rates or relaxation processes, $S_\mathrm{ex}$ is non-zero.
It has been shown that $S_\mathrm{ex}$ is deeply related to
the second law of steady state thermodynamics;
there is the following inequality relation between the excess entropy production $S_\mathrm{ex}$
and the change of the Shannon entropy $\Delta S$:
\begin{eqnarray}
S_\mathrm{ex} \ge - \Delta S.
\label{eq_extended_second_law}
\end{eqnarray}
This inequality is always satisfied even when we consider nonadiabatic changes of the 
nonequilibrium states.

The excess entropy production $S_\mathrm{ex}$ is one of examples of the excess contributions.
As discussed in the above, 
one of characteristics of the excess contribution is as follows:
when the transition rates are fixed,
the excess contribution converges to a certain value after some relaxation time;
it does not grow with time any more.
In contrast, the house keeping contribution gives steadily a non-zero value
even when the system remains in a single nonequilibrium steady states
and there are no changes of the transition rates.
Note that there could be various ways to define the excess contributions.
$S_\mathrm{ex}$ is only one of the examples,
although, of course, $S_\mathrm{ex}$ is an interesting and important one
because it satisfies the important property \eref{eq_extended_second_law}.

Does the contribution from the geometric phase
have the same characteristics with the excess contribution?
If the time-evolution operator $H$ is time-independent,
it is clear that the ket vector $|\psi(\chi,t)\rangle$ finally 
converges to the right eigenstate with the largest eigenvalue;
the bra vector $\langle \psi (\chi,t)|$ also converges
to the left eigenstate.
In addition, $\frac{\rmd}{\rmd t} | \phi(\chi,t) \rangle = 0$ 
when the system is in the eigenstate.
This means that the geometric phase $\gamma(\chi,T)$ converges to a certain value
after some relaxation time.
Hence, we can conclude that the geometric phase has the characteristics of the excess contribution adequately.



\section{Example}

In order to investigate the characteristics of the noncyclic geometric phase,
we here consider a simple hopping model,
which has been used in order to study the geometric contribution under cyclic perturbations
\cite{Sinitsyn2007,Sinitsyn2007a,Ohkubo2008,Ohkubo2008a}:
\begin{eqnarray}
\begin{array}{ccccc}
\multirow{2}{20pt}{$\Bigg[\mathrm{L}\Bigg]$}  & \stackrel{w_{21}^{1}}{\rightarrow} & 
\multirow{2}{60pt}{$\Bigg[\mathrm{container}\Bigg]$}  & \stackrel{w_{12}^{2}}{\rightarrow} & 
\multirow{2}{20pt}{$\Bigg[\mathrm{R}\Bigg]$} \\
& \stackrel{w_{12}^{1}}{\leftarrow} &   & \stackrel{w_{21}^{2}}{\leftarrow} & 
\end{array}.
\end{eqnarray}
The system consists of three parts.
The container can contain at most one particle in it,
and then the container has only two states, i.e., filled state or empty one.
When the container is filled with one particle,
the particle can escape from the container by jumping into one of the two absorbing states:
the left reservoir or the right one.
In contrast, when the container is empty, the left or right reservoirs 
can emit a new particle into the container.
The master equation for the system is written as follows:
\begin{eqnarray}
\fl
\frac{\partial}{\partial t}
\left( \begin{array}{c} p_\mathrm{1}(t) \\ p_\mathrm{2}(t) \end{array} \right)
= 
\left(
\begin{array}{cc}
-w_{21}^{1}(t) - w_{21}^{2}(t) & w_{12}^{1}(t) + w_{12}^{2}(t)  \\
w_{21}^{1}(t) + w_{21}^{2}(t)  & - w_{12}^{1}(t) - w_{12}^{2}(t)
\end{array}
\right)
\left( \begin{array}{c} p_\mathrm{1}(t) \\ p_\mathrm{2}(t) \end{array} \right),
\label{eq_example_master_equation}
\end{eqnarray}
where $p_1(t)$ and $p_2(t) = 1-p_1(t)$ correspond to 
the probabilities with which the container is empty
and filled, respectively.

\subsection{Hopping from the container to the right reservoir}

\begin{figure}
\begin{center}
\includegraphics[width=100mm]{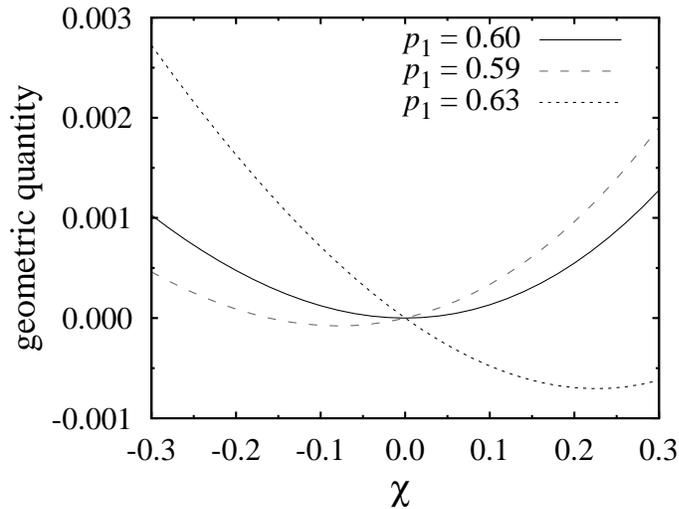}
\end{center}
\caption{
The geometric quantity $\gamma(\chi,T)$.
Results with three different initial probabilities $p_1(t=0)$ are shown.
$p_1 = 0.60$ corresponds to the stationary probability of the system.
}
\label{fig_example}
\end{figure}

Here, we calculate the number of hopping from the container to the right reservoir.
Hence, the quantities $\{\alpha_{ss'}^{\nu}\}$ is selected as follows:
$\alpha_{12}^{2} = 1$ and $\alpha_{ss'}^{\nu} = 0$ for other cases.
The time-evolution operator $H(\chi,t)$ is therefore written as 
\begin{eqnarray}
H(\chi,t) = 
\left(
\begin{array}{cc}
-w_{21}^{1}(t) - w_{21}^{2}(t) & w_{12}^{1}(t) + w_{12}^{2}(t) \rme^{\chi}   \\
w_{21}^{1}(t) + w_{21}^{2}(t)  & - w_{12}^{1}(t) - w_{12}^{2}(t)
\end{array}
\right).
\end{eqnarray}

We consider relaxation processes of the system.
That is, the transition rates $\{w_{ss'}^{\nu}\}$ are assumed to be time-independent,
and we investigate effects of the initial probability $p_1 \equiv p_1(t=0)$.
If the probability $p_1$ is the stationary one for 
the master equation \eref{eq_example_master_equation},
the system is staying in the stationary state.
On the other hand, if the probability $p_1$ is not the stationary one,
the system relaxes to the stationary state.
The non-stationary initial $p_1$ corresponds to a stationary state
with a different transition matrix,
and hence the above situation can be considered
as the case in which there are sudden changes for the transition rates
at time $t=0$ by external forces.

Note that the geometric phase exists even in the stationary initial states.
That is, the initial state is $| p(t=0) \rangle$ for \eref{eq_time_evolution_for_phi_ket},
and $\langle 1 |$ for \eref{eq_time_evolution_for_phi_bra},
and these states are not the right and left eigenstates of $H(\chi,t)$.
Hence, $| \phi(\chi,0) \rangle \neq | \phi(\chi,T) \rangle$ and
$\langle \varphi(\chi,0) | \phi(\chi,T) \rangle \neq 1$ in general.
However, for the stationary initial cases, there should not be any contribution
to the current from the geometric phase;
since the geometric phase corresponds to the cumulant generating function,
the first derivative must vanish for the stationary initial case.
Here, we will check the fact that the geometric phase for the stationary initial case
does not give a current, as expected.

Setting $w_{12}^{2} = 2$ and $w_{ss'}^{\nu} = 1$ for other cases,
we numerically evaluate the geometric phase $\gamma(\chi,T)$.
Here, the final time $T$ is selected large enough for the relaxation.
Figure~\ref{fig_example} shows the results.
When we set the initial probability $p_1(t=0)$ as $0.60$,
which corresponds to the stationary probability of the given transition rates,
the first derivative at $\chi = 0$ is zero.
This means that there is no contribution to the current from the geometric phase.
On the other hand, 
non-stationary initial probabilities $p_1(t=0)$ give
a current due to the geometric phase;
the first derivative at $\chi = 0$ is non-zero.
Note that the geometric phases are calculated for large $T$,
and we have confirmed the convergence of these quantities.
Hence, it is possible to consider the contribution
from the geometric phase as an excess contribution.

We here comment on the higher-order contribution from the geometric phase.
Figure~\ref{fig_example} shows that there is a contribution to the current fluctuation of the system
from the geometric phase, even when the system starts from the steady state.
One may consider that it is strange, because
the current and its statistics in the steady state may be characterized only by
the largest eigenvalue of the time-evolution operator $H(\chi,t)$.
However, it has been known that 
there is another source of the contribution for the current and its statistics;
for example, so-called `boundary terms' explicitly appear when we use the path-integrals.
There are at least two different proposals for the boundary terms \cite{Sinitsyn2010,Sagawa2011};
in one proposal, the boundary terms are included in the dynamical phase \cite{Sinitsyn2010};
in another proposal, they are in the geometric phase \cite{Sagawa2011}.
The formulation based on the fiber bundles suggests a different consequence;
the boundary terms are divided into two parts in some way
and contribute to both the dynamical and geometric phases.

\subsection{Entropy production}

\begin{figure}
\begin{center}
\includegraphics[width=77mm]{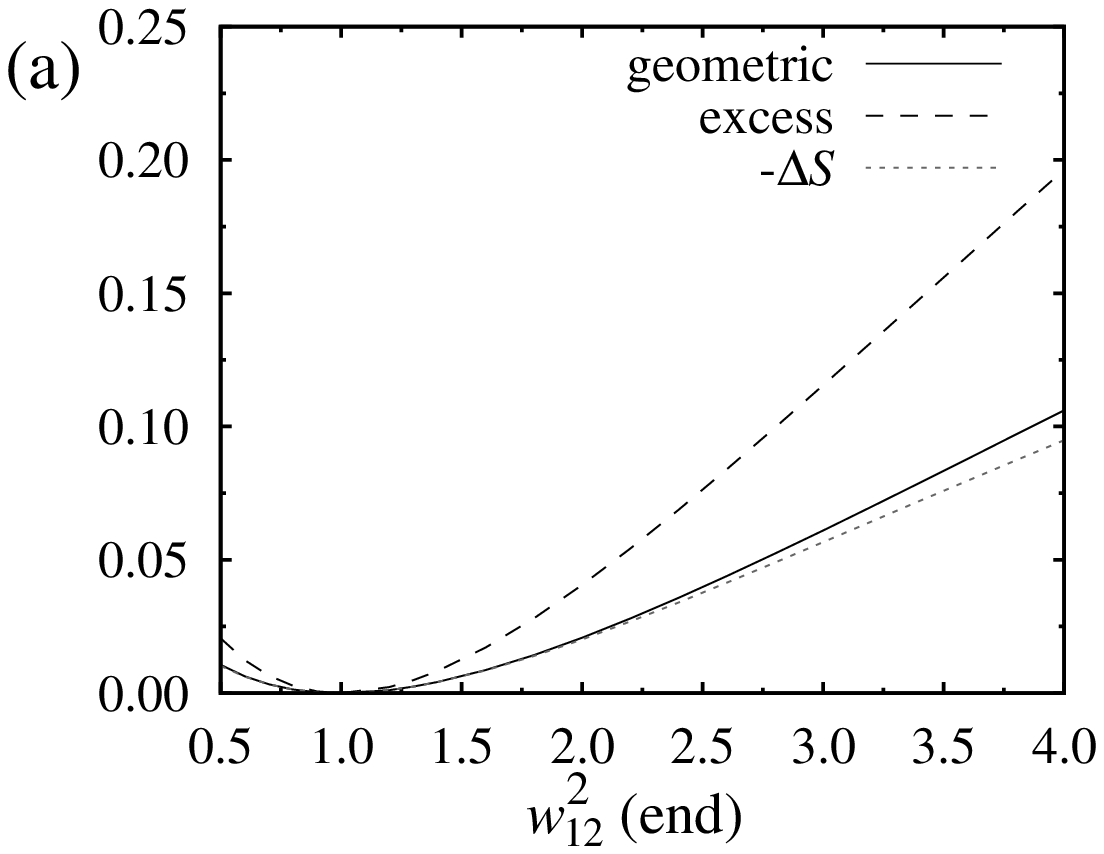}
\includegraphics[width=77mm]{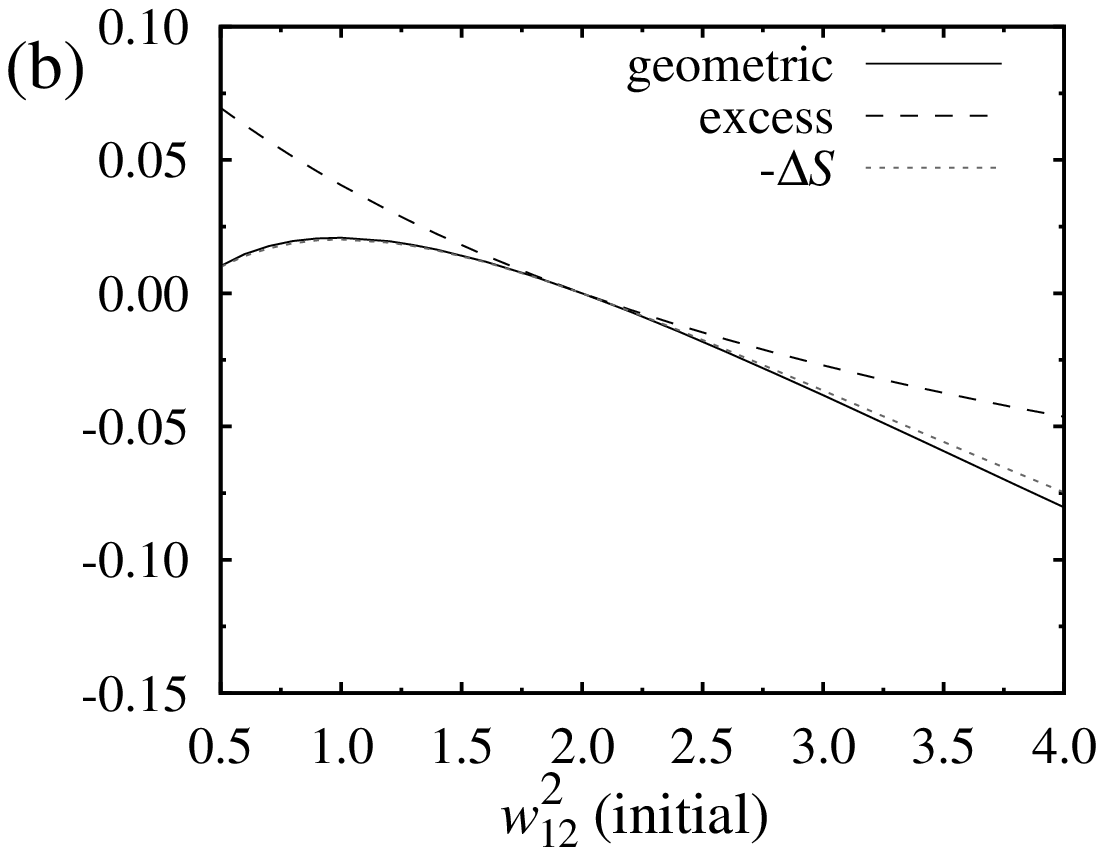}
\end{center}
\caption{
The contribution of the entropy production from the geometric phase,
$S_\mathrm{geom}$, which is evaluated from \eref{eq_d_EP_geom},
and the conventional excess entropy production, $S_\mathrm{ex}$, defined in \eref{eq_excess_EP}.
The changes of the Shannon entropy are also depicted.
(a) Transitions from an equilibrium state to nonequilibrium states.
(b) Transitions between nonequilibrium states.
For both cases, we start from a steady state at $t = 0$,
and one of the transition rates is suddenly changed at $t = 5$.
For (a), all transition rates $w_{ss'}^{\nu} = 1.0$ for $t < 5$, 
and only $w_{12}^{2}$ is changed to a certain value 
($w_{12}^{2}$ (end) in the x-axis) for $t \ge 5$.
For (b), $w_{12}^{2}$ is set to a certain value ($w_{12}^{2}$ (initial) in the x-axis) for $t < 5$,
and $w_{12}^{2} = 2.0$ for $t \ge 5$. All other transition rates are $1.0$ and time-independent.
}
\label{fig_entropy}
\end{figure}

As an example of interesting physical quantities,
we next demonstrate the contribution of the geometric phase to the entropy production,
which are discussed in section 3.3.
The matrix $H(\chi,t)$ for the entropy production is given as
\begin{eqnarray}
\fl
H(\chi,t) = 
\left(
\begin{array}{cc}
-w_{21}^{1}(t) - w_{21}^{2}(t) & w_{12}^{1}(t) \rme^{\chi \alpha_{12}^{1}} + w_{12}^{2}(t) \rme^{\chi \alpha_{12}^{2}}   \\
w_{21}^{1}(t) \rme^{\chi \alpha_{21}^{1}} + w_{21}^{2}(t) \rme^{\chi \alpha_{21}^{2}}  & - w_{12}^{1}(t) - w_{12}^{2}(t)
\end{array}
\right),
\end{eqnarray}
where $\alpha_{ss'}^{\nu}$ is defined as \eref{eq_def_EP}.
In order to discuss the comparison with the excess entropy production in \eref{eq_excess_EP},
the `average' value of the contribution from the geometric phase should be evaluated.
It is easy to calculate the average contribution from the geometric phase
as the subtraction of the contribution from the dynamical phase
from the total contribution:
\begin{eqnarray}
\frac{\rmd}{\rmd t} S_\mathrm{geom}
&= \langle 1 |
\left. \frac{\partial}{\partial \chi} \Big( \frac{\rmd}{\rmd t} | \psi(\chi,t) \rangle \Big) \right|_{\chi=0}
- \left. \frac{\partial}{\partial \chi} \delta(\chi,t) \right|_{\chi=0}
\nonumber \\
&= - \left. \frac{\partial}{\partial \chi} \Big( \langle \psi(\chi,t)| \Big) \right|_{\chi=0}
H(\chi=0,t) | p(t) \rangle,
\label{eq_d_EP_geom}
\end{eqnarray}
and its integral gives the contribution $S_\mathrm{geom}$.
In order to evaluate \eref{eq_d_EP_geom},
it is needed to calculate the time evolutions of the derivative of the bra vector with respect to $\chi$,
$\partial \langle \psi(\chi,t)| / \partial \chi |_{\chi = 0}$,
and the time evolution equation is easily obtained from \eref{eq_time_evolution_for_psi_bra}.
Of course, the solutions of the original master equation, $|p(t)\rangle$, are also needed.

Note that the definitions of $S_\mathrm{geom}$ and $S_\mathrm{ex}$ are completely different,
and actually, the characteristics are also different.
For example, in the counting statistics,
the initial value for the bra vector $\langle \psi(\chi,t) |$ is $\langle 1 |$,
and then its derivative with respect to $\chi$ is zero.
Hence, the time-derivative \eref{eq_d_EP_geom} is zero for the initial value.
On the other hand, the time-derivative of the excess entropy production in \eref{eq_excess_EP} 
has a finite value for the initial state in general.

Here, in order to avoid the boundary term problems discussed in section 4.1, 
the following settings are used;
we start from $t = 0$, and for $t < 5$, the transition rates are time-independent.
At $t = 5$, one of the transition rates, $w_{12}^{2}$, is suddenly changed,
and we investigate the relaxation process.
We confirmed that the initial time interval ($t = 0 \sim 5$) is enough to eliminate
the effects of the boundary term problem.

Firstly, we study a case with transitions from an equilibrium state to nonequilibrium states.
That is, for $t < 5$, all transition rates $w_{ss'}^{\nu}$ is fixed to $1.0$,
and for $t \ge 5$, $w_{12}^{2}$ is changed to a certain value (the x-axis in Figure~\ref{fig_entropy}(a)).
All quantities are numerically evaluated 
for enough large $T$ ($T = 10$).
The results are shown in Figure~\ref{fig_entropy}(a).
The excess entropy production is always larger than the change of the Shannon entropy,
as expected from \eref{eq_extended_second_law}.
While the contribution from the geometric phase seems also be always larger than $- \Delta S$,
there is no guarantee to satisfy the inequality relation, as we will see soon.

Secondly, transitions between nonequilibrium states are investigated.
In this case, for $t < 5$, $w_{12}^{2}$ is set to a certain value (the x-axis in Figure~\ref{fig_entropy}(b)).
And for $t \ge 5$, $w_{12}^{2}$ is changed to $2.0$.
All other transition rates are $1.0$ and time-independent.
In this case, it is clear that the contribution from the geometric phase
can be smaller than $-\Delta S$.

Different from $S_\mathrm{ex}$,
$S_\mathrm{geom}$ would not simply connected to the inequality with the change of the Shannon entropy,
In this sense, one may consider that 
$S_\mathrm{geom}$ could not be useful to construct steady-state thermodynamics.
On the other hand, $S_\mathrm{geom}$ is closer to $-\Delta S$ than $S_\mathrm{ex}$
for a wide range of parameter changes.
Although one may expect that $S_\mathrm{geom}$ gives $- \Delta S$ precisely 
because they are quite similar in Figures~\ref{fig_entropy}(a) and (b),
as far as we checked, there are the differences between $S_\mathrm{geom}$ and $- \Delta S$.
In order to connect the geometric phase to physical quantities,
more studies for mathematical frameworks and physical examples will be needed.
At least, when we consider a small change of transition rates,
$S_\mathrm{geom}$ may be more useful compared with $S_\mathrm{ex}$;
the connections of the geometric phase with the Shannon entropy or other informational quantities
should be investigated in future.

\section{Concluding remarks}

In the present paper, we proposed a natural division of the cumulant generating function
for the current.
The discussion is based on the fiber bundles and
we confirmed that the geometric phase has the characteristics of excess contributions.
In addition, as a physical example, 
some discussions for entropy production were given.
As stated before, at this stage, there are only several applications
of the geometric phase to nonequilibrium physics.
However, there is a mathematical background of the fiber bundles,
and the framework is applicable to various physical quantities.
We hope that the present paper motivates future works
of applications of geometric phases and fiber bundles
to nonequilibrium physics.

We finally comment on some unclear points.
As discussed in section 4,
it has been revealed that even in the steady state
the geometric phase has the contribution to the fluctuations.
The physical meaning of this fact should be investigated in future works
although the contribution from the geometric phase will be very small
comparing to that from the dynamical phase in general.
In addition, mathematically more rigorous and extended discussions may be also needed;
the non-Hermite property could cause difficulty for the settings of the fiber bundle and its dual.
A cyclic time-evolution for non-Hermitian cases has been discussed in \cite{Garrison1988},
and the settings in the present paper seem to adequately correspond to the formulation in \cite{Garrison1988}.
However, there may be more adequate discussions for the non-Hermite cases,
which might not need the geodesic curves or dual structures in order to define geometric phases;
such different mathematical formulations may enable us to give extended discussions.



\section*{Acknowledgments}
This work was supported in part by grant-in-aid for scientific research 
(Grant Nos.~20115009 and 25870339)
from the Ministry of Education, Culture, Sports, Science and Technology (MEXT), Japan.

\appendix
\section{Comment on the geodesic curve}

Consider the trajectory in $P$, which is made by the time-evolution from the initial ket vector
$| \phi(\chi,t=0) \rangle$ to the final ket vector $| \phi(\chi,t=T) \rangle$.
In order to make a closed curve in $P$, a concept of `geodesic' curves is useful \cite{Samuel1988}.

Firstly, we explain how to derive the geodesic curve.
Considering a curve in $P$, $|\phi(\chi,s)\rangle$,
we introduce the tangent vector to this curve as $| u\rangle = \frac{\rmd}{\rmd s} |\phi(\chi,s)\rangle$,
as discussed in section 2.
In addition, we also introduce the corresponding tangent vector in the dual space as
$\langle v | = \frac{\rmd}{\rmd s} \langle \varphi (\chi,s) |$.
In the following discussions, 
we will use $A^{\mathrm{d}}(\chi,s)$ in order to express the connection
in the dual space,
but we assume that the dual curve is selected so that the inner produce is conserved;
i.e., $A(\chi,s) = - A^{\mathrm{d}}(\chi,s)$ as discussed in section 2.
Under a gauge transformation, the tangent vectors $|u\rangle$ and $\langle v|$
are not gauge covariant, and therefore the following gauge covariant quantities are useful:
\begin{eqnarray}
| u'(\chi,s) \rangle &\equiv 
\frac{\rmd}{\rmd s} | \phi(\chi,s) \rangle - A(\chi,s) | \phi(\chi,s) \rangle, \\
\langle v'(\chi,s) | &\equiv
\frac{\rmd}{\rmd s} \langle \varphi(\chi,s) | - \langle \varphi(\chi,s)| A^{\mathrm{d}}(\chi,s),
\end{eqnarray}
Hence, $\langle v' | u' \rangle$ is gauge invariant,
and  $\langle v' | u' \rangle$ is available in order to define a kind of metric. 
The variation of $\int \langle v' | u' \rangle \rmd l$,
where $l$ is an affine parameter,
gives the geodesic curves $| \phi(\chi,s) \rangle$ and $\langle \phi(\chi,s)|$, which are determined by
\begin{eqnarray}
&\frac{\rmd}{\rmd s} | u'(\chi,s) \rangle - A(\chi,s) | u'(\chi,s) \rangle = 0, \label{eq_geodesic_1} \\
&\frac{\rmd}{\rmd s} \langle v'(\chi,s) | - \langle v'(\chi,s) | A^{\mathrm{d}}(\chi,s) = 0. \label{eq_geodesic_2}
\end{eqnarray}

Using the geodesic curve,
it is possible to evaluate the integral \eref{eq_integral_of_A_result}.
We consider a geodesic curve $| \phi^\mathrm{g}(\chi,q)\rangle$ ($q \in [0,1]$)
connecting to the trajectory obtained from the time-evolution backward;
$|\phi^\mathrm{g}(\chi,q=0)\rangle = |\phi(\chi,t=T)\rangle$
and $|\phi^\mathrm{g}(\chi,q=1)\rangle = |\phi(\chi,t=0)\rangle$.
Hence, we have a closed curve in $P$,
i.e.,
the actual time-evolution from $|\phi(\chi,t=0)\rangle$ to $|\phi(\chi,t=T)\rangle$,
and back along the geodesic curve $| \phi^\mathrm{g}(\chi,q)\rangle$.
We can therefore define the integral $\oint A(\chi,s) \rmd s$ along the closed curve.
Note that $A(\chi,t) = 0$ on the actual time-evolution,
and $A(\chi,s) \neq 0$ on $| \phi^\mathrm{g}(\chi,q)\rangle$ in general.

In order to calculate the value of the integral,
the characteristics of the geodesic curve is important.
Using a certain gauge transformation on 
$| \phi^\mathrm{g}(\chi,q)\rangle = \exp(-\beta(q)) | \tilde{\phi}^\mathrm{g}(\chi,q)\rangle$,
where $\beta(q=0) = 0$,
we choose a specific connection $\tilde{A}(\chi,q) = 0$
on $| \tilde{\phi}^\mathrm{g}(\chi,q)\rangle$.
Due to the gauge transformation, the derived whole curve, including the actual time-evolution, is not closed.
Note that \eref{eq_geodesic_1} and \eref{eq_geodesic_2} are gauge covariant,
and the gauge transformed curve $| \tilde{\phi}^{\mathrm{g}}(\chi,q)\rangle$ is still a geodesic curve.
In order to have a closed curve, we must use a vertical curve
joining $| \tilde{\phi}^\mathrm{g}(\chi,q=1)\rangle$ and 
$| \phi(\chi,t=0)\rangle$.
It is clear that the vertical contribution corresponds to the integral $\oint A(\chi,s) \rmd s$,
because $A(\chi,t)=0$ on the actual time-evolution and 
$\tilde{A}(\chi,s) = 0$ on the gauge transformed curve.
In addition, the contribution from the vertical curve is given by $\beta(q=1)$.

In order to evaluate $\beta(q=1)$, 
the quantity $g(q) = \langle \tilde{\varphi}^{\mathrm{g}}(\chi,q)| \phi(\chi,t=T) \rangle$,
where $\langle \tilde{\varphi}^{\mathrm{g}}(\chi,q)|$ is the dual of $| \tilde{\phi}^{\mathrm{g}}(\chi,q)\rangle$,
is useful.
Then, we have $g(0) = 1$ because 
$\langle \tilde{\varphi}^{\mathrm{g}}(\chi,q=0)| \phi(\chi,t=T) \rangle
= \langle \varphi^{\mathrm{g}}(\chi,q=0)| \phi(\chi,t=T) \rangle = 1$;
$\frac{\rmd}{\rmd q} g(0) = 0$, which is obtained from $\tilde{A}(\chi,q=0) = 0$;
and $\frac{\rmd^2}{\rmd q^2} g(q) = 0$,
which stems from the definitions of the geodesic curve.
These facts suggest that $g(q)$ is constant for all $q$.
Hence, 
\begin{eqnarray}
1 
&= \langle \varphi^{\mathrm{g}}(\chi,q=0)| \phi(\chi,t=T) \rangle 
= \langle \tilde{\varphi}^{\mathrm{g}}(\chi,q=0)| \phi(\chi,t=T) \rangle \nonumber \\
&= \langle \tilde{\varphi}^{\mathrm{g}}(\chi,q=1) |  \phi(\chi,t=T)\rangle 
=  \rme^{-\beta(q=1)} \langle \varphi^{\mathrm{g}}(\chi,q=1) | \phi(\chi,t=T) \rangle \nonumber \\
&=  \rme^{-\beta(q=1)} \langle \varphi(\chi,t=0) | \phi(\chi,t=T) \rangle.
\end{eqnarray}
That is, the vertical contribution is evaluated as 
$\beta(q=1) = \ln \langle \varphi(\chi,t=0)| \phi(\chi,t=T) \rangle$,
which gives \eref{eq_integral_of_A_result}.


\section*{References}
\begin{thebibliography}{99}


\bibitem{Oono1998} Oono Y and Paniconi M 1998 {\it Prog. Theor. Phys. Supple.} {\bf 130} 29
\bibitem{Hatano2001} Hatano T and Sasa S 2001 {\it Phys. Rev. Lett.} {\bf 86} 3463
\bibitem{Speck2005} Speck T and Seifert U 2005 {\it J. Phys. A: Math. Gen.} {\bf 38} L581
\bibitem{Esposito2007} Esposito M, Harbola U and Mukamel S 2007 {\it Phys. Rev. E} {\bf 76} 031132
\bibitem{Komatsu2008} Komatsu T S and Nakagawa N 2008 {\it Phys. Rev. Lett.} {\bf 100} 030601
\bibitem{Komatsu2008a} Komatsu T S, Nakagawa N, Sasa S I and Tasaki H 2008
{\it Phys. Rev. Lett.} {\bf 100} 230602
\bibitem{Komatsu2009} Komatsu T S, Nakagawa N, Sasa S I and Tasaki H 2009 
{\it J. Stat. Phys.} {\bf 134} 401
\bibitem{Esposito2010} Esposito M and den Broeck C V 2010 {\it Phys. Rev. Lett.} {\bf 104} 090601

\bibitem{Sinitsyn2007} Sinitsyn N A and Nemenman I 2007 {\it Europhys. Lett.} {\bf 77} 58001
\bibitem{Sinitsyn2007a} Sinitsyn N A and Nemenman I 2007 {\it Phys. Rev. Lett.} {\bf 99} 220408
\bibitem{Ohkubo2008} Ohkubo J 2008 {\it J. Stat. Mech.} P02011
\bibitem{Ohkubo2008a} Ohkubo J 2008 {\it J. Chem. Phys.} {\bf 129} 205102
\bibitem{Sinitsyn2008} Sinitsyn N A and Saxena A 2008 {\it J. Phys. A: Math. Theor.} {\bf 41} 392002
\bibitem{Sinitsyn2009} Sinitsyn N A 2009 {\it J. Phys. A: Math. Theor.} {\bf 42} 193001



\bibitem{Sinitsyn2010} Sinitisyn N A and Nemenman I 2010 {\it IET Syst. Biol.} {\bf 4} 409
\bibitem{Sagawa2011} Sagawa T and Hayakawa H 2011 {\it Phys.Rev. E} {\bf 84} 051110

\bibitem{Ohkubo2010} Ohkubo J and Eggel T 2010 {\it J. Phys. A: Math. Theor.} {\bf 43} 425001

\bibitem{Gopich2006} Gopich I V and Szabo A 2006 {\it J. Chem. Phys.} {\bf 124} 154712
\bibitem{Garrahan2009}
Garrahan J P, Jack R L, Lecomte V, Pitard E, van Duijvendijk K and van Wijland F 2009
{\it J. Phys. A: Math. Theor.} {\bf 42} 075007


\bibitem{Bohm_book} Bohm A, Mostafazadeh A, Koizumi H, Niu Q and Zwanziger J
2003 {\it The Geometric Phase in Quantum Systems} (Berlin: Springer-Verlag)

\bibitem{Nakahara_book} Nakahara M 2003 {\it Geometry, Topology, and Physics 2nd ed.}
(Boca Raton: Taylor \& Francis)


\bibitem{Samuel1988} Samuel J and Bhandari R 1988 {\it Phys. Rev. Lett.} {\bf 60} 2339

\bibitem{Garrison1988} Garrison J C and Wright E M 1988 {\it Phys. Lett. A} {\bf 128} 177




\end{thebibliography}
\end{document}